\documentclass[letter]{aa}
\usepackage{natbib}
\usepackage{graphicx}

\usepackage{txfonts}
\begin{document}

   \title{Ross 128 - GL 447}

   \subtitle{A possible activity cycle for a slow-rotating fully-convective star.}

   \author{R.V. Iba\~nez Bustos
          \inst{1,2}\fnmsep\thanks{Based on data obtained at Complejo Astron\'omico El Leoncito, operated under agreement between the Consejo Nacional de Investigaciones Cient\'\i ficas y T\'ecnicas de la Rep\'ublica Argentina and the National Universities of La Plata, C\'ordoba and San Juan.}
          \and
          A.P. Buccino\inst{1,3}
          \and
          M. Flores\inst{4,5}
          \and
          P.J.D. Mauas\inst{1,3}
          }

   \institute{Instituto de Astronomía y Física del Espacio (CONICET-UBA), C.C. 67 Sucursal 28, C1428EHA-Buenos Aires, Argentina.\\
        \and
             Departamento de Física. FI-Universidad de Buenos Aires, Buenos Aires, Argentina.\\
        \and
             Departamento de Física. FCEyN-Universidad de Buenos Aires, Buenos Aires, Argentina.\\
        \and
            Instituto de Ciencias Astronómicas, de la Tierra y del Espacio (ICATE-CONICET), San Juan, Argentina.\\
        \and
            Facultad de Ciencias Exactas, Físicas y Naturales, Universidad Nacional de San Juan, San Juan, Argentina.
             }

   \date{}

 
  \abstract
   {Long-term chromospheric activity in slow-rotating fully-convective stars has been scarcely explored. Ross 128 (Gl 447) is a slow-rotator and inactive dM4 star which has been extensively observed. It hosts the fourth closest extrasolar planet. }
   {Ross 128 is an ideal target to test dynamo theories in slow-rotating low-mass star.}
   { To  characterize the magnetic activity of Ross 128 we study the $S_K$-indexes derived from CASLEO, HARPS, FEROS, UVES and XSHOOTER spectra. Using the Generalized Lomb-Scargle and CLEAN periodograms, we analyze the whole $S_K$ time-series obtained between 2004 and 2018. We perform a similar analysis  for the  Na \scriptsize{I}\normalsize\- index and we analyze its relation with the $S_K$-index. }
   {From both  indexes, we obtain a possible activity cycle with a $\sim$5-year period, which is one of a small handful of activity cycles reported for a slow-rotating fully convective star.} 
   {}

   \keywords{stars: activity --
                stars: late-type --
                techniques: spectroscopic
               }

   \maketitle
%

\section{Introduction}

Cool main-sequence stars later than M3.5-4V (with masses lower than $\sim$0.35 M$_\odot$) are thought to be fully convective \citep{Chabrier97}. Thus, for decades many authors believed that a standard solar-$\alpha\Omega$ dynamo could not operate in these stars due to the absence of a tachocline (\citealt{Chabrier06,Browning08,Reiners10}). In this context, \cite{Pizzolato03} and \cite{Wright11} studied the X-ray rotation-activity relationship  in  late-type stars through $L_X/L_{bol}$ vs. $P_{rot}$ and $L_X/L_{bol}$ vs. $R_o$ diagrams. In both diagrams they found two regimes: saturated and non-saturated. , \cite{Wright11} suggested that this bimodality could be caused by two different dynamos: a convective dynamo parameterized by the bolometric luminosity for fast-rotator stars and an interface-type dynamo parameterized by the Rossby number for slow rotators (P$_{rot}$ $>$ 10 days). However, it was still unclear why there should exist two different dynamo configurations in response to two regimes whose transition seems to be rather smooth. Recently, \cite{Wright18} extended this analysis including slow-rotating fully-convective stars in the $L_X/L_{bol}$ diagram. They concluded that the rotation-activity relationship for fully-convective stars (both slow and fast-rotators) present a similar behavior to partly-convective stars, as also observed in $log(R'_{HK})$ \citep{Astudillo17}. Moreover, multi-wavelength observations of slow rotator stars confirmed that they have strong magnetic fields (\citealt{Hawley91, Reiners09}). Three-dimensional dynamo simulations without a tachocline (\citealt{Browning08, Yadav15, Yadav16}) also show that it is possible to produce large-scale magnetic fields in stellar convection layers. All these studies suggests that an $\alpha\Omega$ dynamo can operate for both, slow and fast-rotator fully-convective stars even without a tachocline.Therefore, the detection of activity cycles in fully-convective stars could help to understand better the different dynamos at work.

On the other hand, in the last decade M stars have become favorable targets to search for Earth-like planets in the solar neighborhood, due to their low-mass and the appreciable star-planet contrast ratio. Around 200 exoplanets have been detected orbiting M dwarf stars through both transit and radial velocity (RV) methods (\citealt{Bonfils05, Udry07, Mayor09, Anglada13, Anglada16, Crossfield15, Gillon17, Dittmann17, Bonfils18, Diaz19}). However, it is well known that different magnetic phenomena as starspots, plage, and activity cycles can induce RV shifts disturbing the detectability of planetary signals \citep{Dumusque14}. The detection of activity cycles in exostars allows us to distinguish between stellar and planetary signals (\citealt{Robertson13, Carolo14, Flores18}). Also, several theoretical studies adressed the impact of stellar activity on planetary atmospheres and habitability (\citealt{Cuntz00, Ip04, Buccino06, Buccino07, Cohen10, Dumusque17}). M dwarfs stars, with particularly strong flares, are ideal to study these impacts.

In the last three years, Earth-like planets orbiting late-type M dwarfs have been reported: Proxima Centauri (M5.5Ve; \citealt{Anglada16}), TRAPPIST-1 (M8; \citealt{Gillon17}), LHS 1140 (M4.5V; \citealt{Dittmann17}) and Ross 128 (M4V; \citealt{Bonfils18}). In particular, the last three stars are inactive (i.e. they do not have H$\alpha$ in emission). In addition, Prox Cen, LHS 1140 and Ross 128 are slow rotators. Of these, only in Proxima Centauri activity cycles have been detected (\citealt{Cincunegui07b, SuarezMascareno16, Wargelin17}). 

To determine the long-term chromospheric activity cycles of late stars, in 1999 we have developed the HK$\alpha$ Project dedicated to periodically obtain mid-resolution echelle spectra of fully and partly-convective stars. Throughout these twenty years, we have found evidence of cyclic activity in M stars under different regimes: the \textit{fully-convective} star Proxima Centauri, (\citealt{Cincunegui07b}), \textit{partly-convective} stars (GJ 229 A and GJ 752 A, \citealt{Buccino11}; AU Mic, \citealt{Ibanez18}) and in the \textit{convective threshold} (the binary system GJ 375, \citealt{Diaz07}; AD Leo, \citealt{Buccino14}). We also found activity cycles in RS CVn \citep{Buccino09} and in the K star $\epsilon$ Eri \citep{Metcalfe13} and $\iota$ Hor \citep{Flores17}.

In this paper we present the  long-term activity study of Ross 128, which has been observed for more than a decade by the HK$\alpha$ Project. 
This star is a fully-convective (M4V) flare star \citep{Lee72} with mass $\sim$ 0.168 M$_{\odot}$ and radius $\sim$ 0.1967 R$_{\odot}$. It is a slow-rotator with a $\sim$ 121 days-period. \cite{Bonfils18} reported a planet orbiting Ross 128 with an orbital period of 9.9 days, which, due to its proximity to the Sun (3.4 parsec), is the fourth closest extrasolar planet. This star is an ideal target to study both the magnetic dynamo in slow-rotating low-mass stars and the impact of stellar activity on planetary atmospheres.
We organized this work as follows: in Section $\S$2 we present our spectroscopic CASLEO observations, which are complemented by public spectroscopic data obtained from the European Southern Observatory (ESO). In Section $\S$3 we analyse the decadal time-series in different activity indicators, and finally we outline our conclusions in Section $\S$4. 


\section{Observations}\label{sec.obs}

In 1999 we started the HK$\alpha$ Project to study long-term chromospheric activity in late-type stars (from dF5 to dM5.5). Our database has currently more than 6000 mid-resolution echelle spectra ($R\approx 13.000$) covering a  wavelength range between 3800 and 6900 \AA. These echelle spectra were obtained with the REOSC\footnote{http://www.casleo.gov.ar/instrumental/js-reosc.php} spectrograph mounted at the 2.15 m Jorge Sahade telescope at the Complejo Astronómico el Leoncito Observatory (CASLEO), in San Juan, Argentina. They were optimally extracted and flux calibrated using IRAF\footnote{The Image Reduction and Analysis Facility (IRAF) is distributed by the Association of Universities for Research in Astronomy (AURA),  Inc., under contract to the National Science Foundation} procedures (see \citealt{Cincu04}, for details).

\begin{table}
\centering
\caption{CASLEO spectra for Ross 128 from the HK$\alpha$ project. Column 1:
  date of each observing run (MMYY). Column 2: Julian date xJD =
  JD - 2450000. Column 3: Exposure time in seconds.} 
\begin{tabular}{ccc}
\hline\hline\noalign{\smallskip}
Label	&	xJD &  t(s)\\
\hline\noalign{\smallskip}
0307 & 4165 & 7200 \\
0308 & 4550 & 7200  \\
0309 & 4904 & 7200  \\
0609 & 4990 & 5400 \\
0612 & 6089 & 5400 \\
0313 & 6353 & 7200 \\
0513 & 6435 & 7260  \\
0614 & 6834 & 5400 \\
0315 & 7109 & 5400 \\
0315 & 7110 & 6000 \\
0316 & 7468 & 5400 \\
0417 & 7858 & 6000 \\ 
0318 & 8203 & 6000 \\
\hline 
\end{tabular}
\label{obs_casleo}
\end{table}

\begin{table}
\caption{ID programs of the Ross 128 observations used in this work.} 
\centering
\resizebox{\hsize}{!}{
\begin{tabular}{l l c c c}
\hline\hline
\multicolumn{2}{c}{HARPS} & FEROS & UVES & XSHOOTER\\
\hline
072.C-0488(E) &   191.C-0873(D)      & 074.D-0016(A) & 091.D-0296(A) & 084.D-0795(A)\\
183.C-0437(A) &   191.C-0873(E)      & 090.A-9003(A) &               & 098.D-0222(A)\\
191.C-0873(A) &   191.C-0873(F)      &               &               &              \\    
\hline 
\end{tabular}
}
\label{obs_HFUX}
\end{table}

In Table \ref{obs_casleo} we show the observation logs of Ross 128 at CASLEO: 13 individual observations distributed between 2007 and 2018. The first column shows the date (month and year) of the observation, the second column lists $xJD = JD - 2450000$, where JD is the Julian date at the beginning of the observation and the third column indicates the observing exposure in seconds.

We complement our data with public observations obtained with several ESO spectrographs, by the Programs listed in Table \ref{obs_HFUX}. 41 spectra were observed by HARPS, mounted at the 3.6 m telescope (\textit{R} $\sim$ 115.000), during two time intervals, $2004-2006$ and $2012-2016$; 3  spectra were obtained in 2005 and 2013 by FEROS, placed on the 2.2 m telescope ($R \sim 48.000$); 3  spectra were taken in 2013 with UVES, attached to the Unit Telescope 2 (UT2) of the Very Large Telescope (VLT) ($R \sim 80.000$); and 11 mid-resolution spectra (\textit{R} $\sim$ 8.900) were obtained in 2014 and 2017 with XSHOOTER, mounted at the UT2 Cassegrain focus, in the UVB wavelength range (300 - 559.5 nm). HARPS and FEROS spectra have been automatically processed by the respective pipelines\footnote{http://www.eso.org/sci/facilities/lasilla/instruments/harps.html}$^,$\footnote{http://www.eso.org/sci/facilities/lasilla/instruments/feros.html}. UVES and XSHOOTER data were reduced with the corresponding procedure\footnote{http://www.eso.org/sci/facilities/paranal/instruments/uves.html}$^,$\footnote{http://www.eso.org/sci/facilities/paranal/instruments/xshooter.html}.

\section{Results}\label{sec.res}

\subsection{The $S_K$-index}

\begin{figure}
\centering
   \includegraphics[width=9cm]{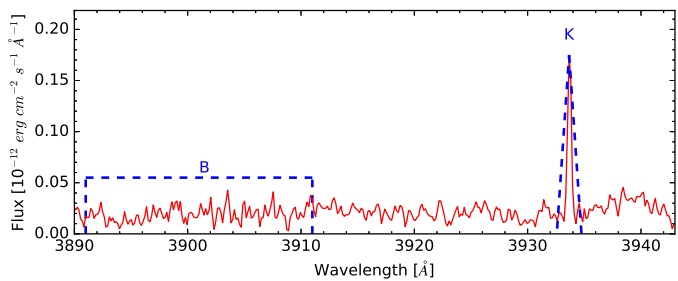}
      \caption{The K line and the blue window for a Ross 128 CASLEO spectrum.
              }
     \label{Sk-index}
\end{figure}

In order to join our dataset with public spectra, which are not calibrated in flux, we need to use a dimensionless index for all the spectra. For decades, the typical stellar activity indicator used for dF to dK stars has been the Mount Wilson $S$-index, defined as the ratio of the Ca \scriptsize{II}\normalsize\- H and K line-core fluxes to the continuum nearby \citep{Baliunas98}. However, as explained in detail in \cite{Buccino11}, it is not the best index to study the chromospheric activity of the dM4 star Ross 128 because in our CASLEO spectra the Ca \scriptsize{II}\normalsize\- K line presents greater signal-to-noise ratio than the Ca \scriptsize{II}\normalsize\- H line. In this work, we use as a proxy of stellar activity the $S_K$-index defined as the ratio between the Ca \scriptsize{II}\normalsize\- K line-core flux integrated with a triangular profile of 1.09 \AA\- FWHM centered in 3933.66 \AA\- and the blue  20 \AA\- window centered at 3901 \AA\- (see Fig. \ref{Sk-index}).

To corroborate the accuracy of the $S_K$-index proposed, we computed this index for  M dwarf stars previously studied by our group.
We applied a Bayesian analysis to evaluate  its correlation with the chromospheric activity indexes reported in \cite{Buccino11}, \cite{Buccino14} and \cite{Ibanez18}.  Using the \textsf{python} code provided by \cite{Figueira16}, we estimated the posterior probability distribution of the correlation coefficient $\rho$. In all cases, we obtained positive correlation coefficients larger than  0.75\footnote{$\rho= (0.750 \pm 0.105)$ for Gl 221A,  $\rho= (0.9467 \pm 0.028)$ for Gl 752A, $\rho= (0.859 \pm 0.073)$ for AD Leo and $\rho= (0.825 \pm 0.041)$ for AU Mic}  whith a 95\% confidence level.

\begin{figure}
\centering
    \includegraphics[width=7.1cm]{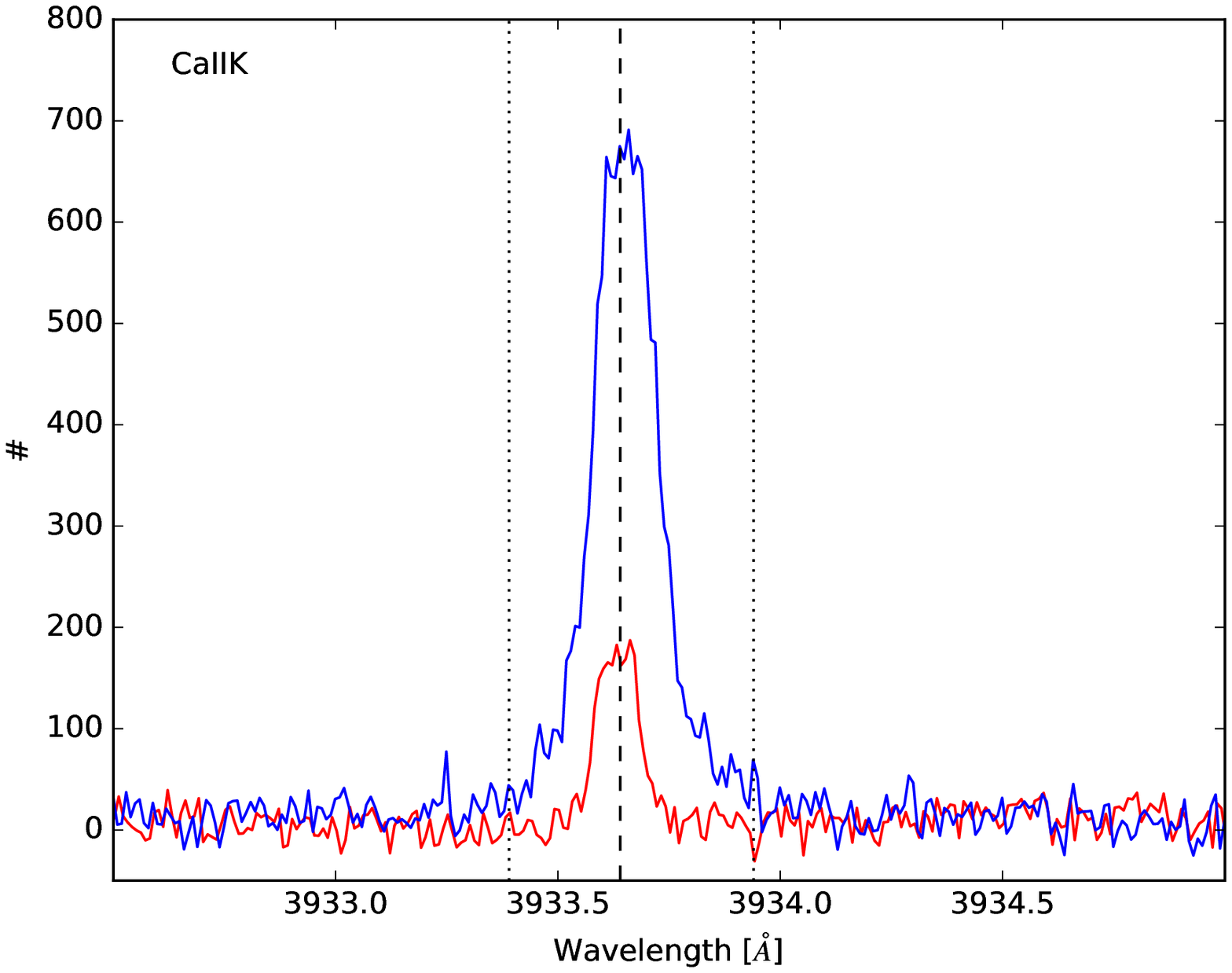}\\
    \includegraphics[width=7.5cm]{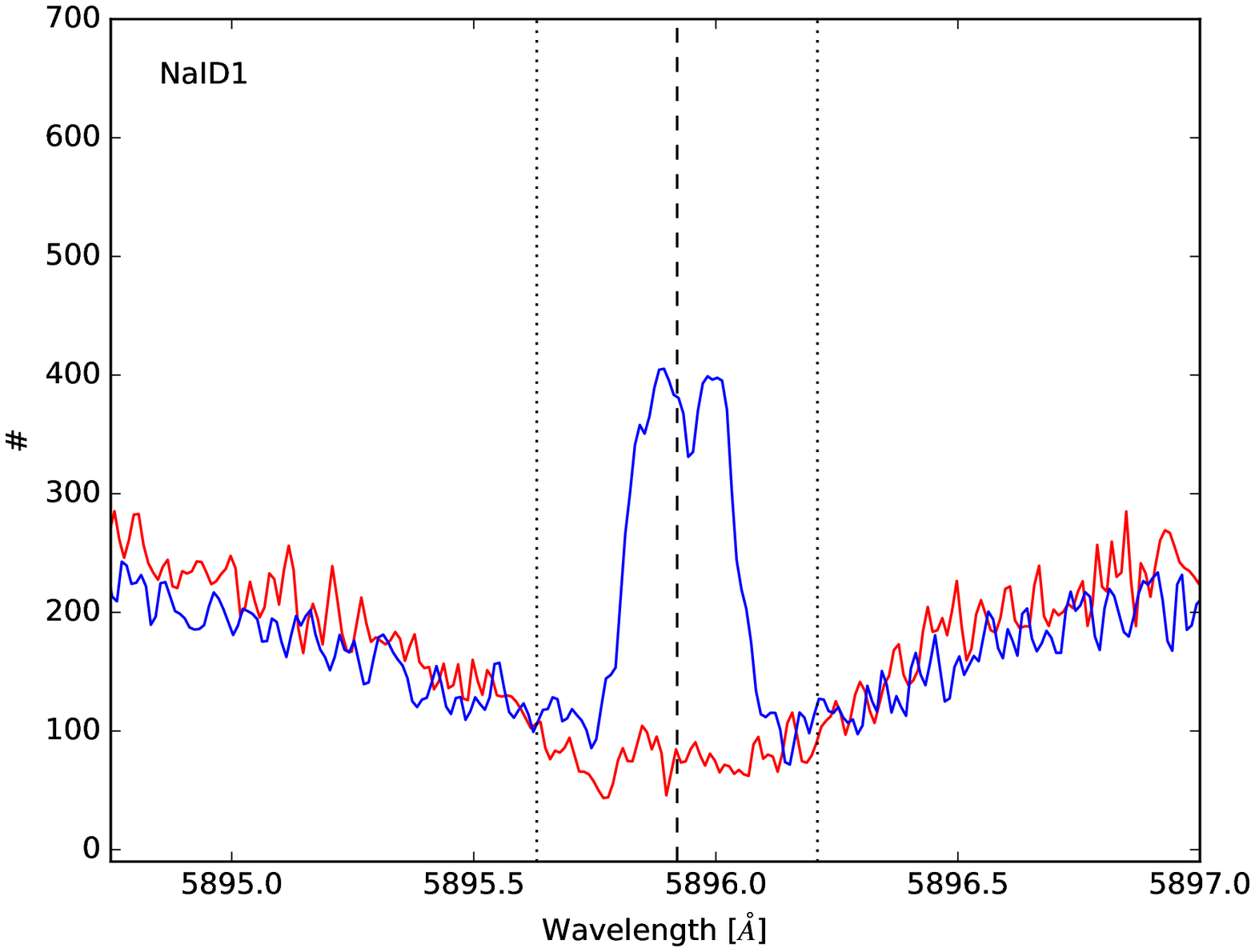}
    \includegraphics[width=7.5cm]{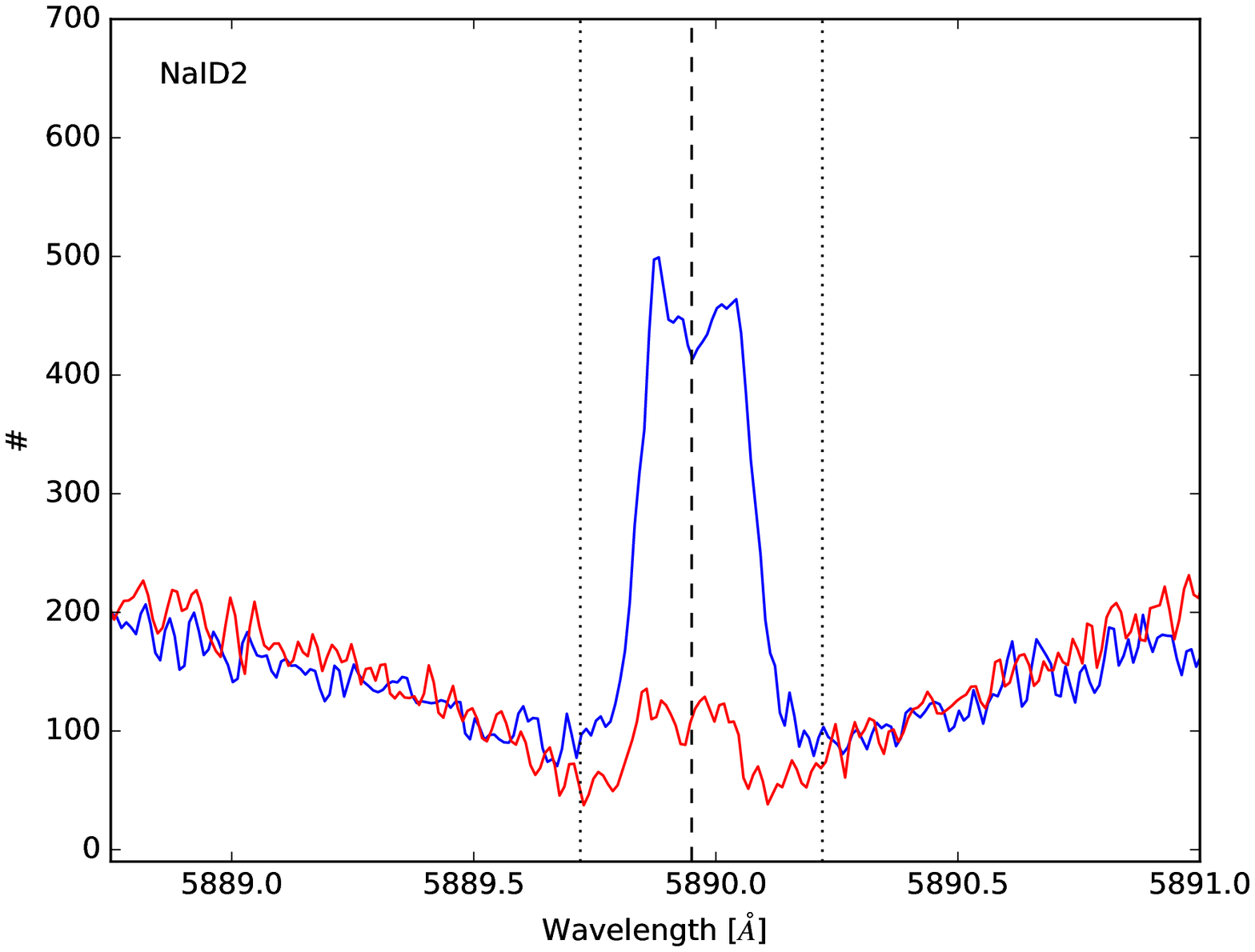}
       \caption{Ca \scriptsize{II}\normalsize\- K, Na I D1 and D2 chromospheric lines at different activity levels: flare (blue) and non-flare spectrum (red).}%
   \label{indices}
\end{figure}

Before exploring the existence of an activity cycle in this star, we filter out any flares from the observations. We visually compare the spectra of nights separated by less than a week, and we discard those with Ca \scriptsize{II}\normalsize\- K flux larger than 2$\sigma$ of the mean value. An example can be seen in Fig.\ref{indices}, were we show the Ca \scriptsize{II}\normalsize\- K, Na \scriptsize{I}\normalsize\- D1 and the Na \scriptsize{I}\normalsize\- D2 lines for two close nights with different activity levels. It can be seen that the flare is not only present in the calcium line, but also in the Na lines. 

\subsection{Long-term activity of Ross 128}

\begin{figure*}
\centering
   \includegraphics{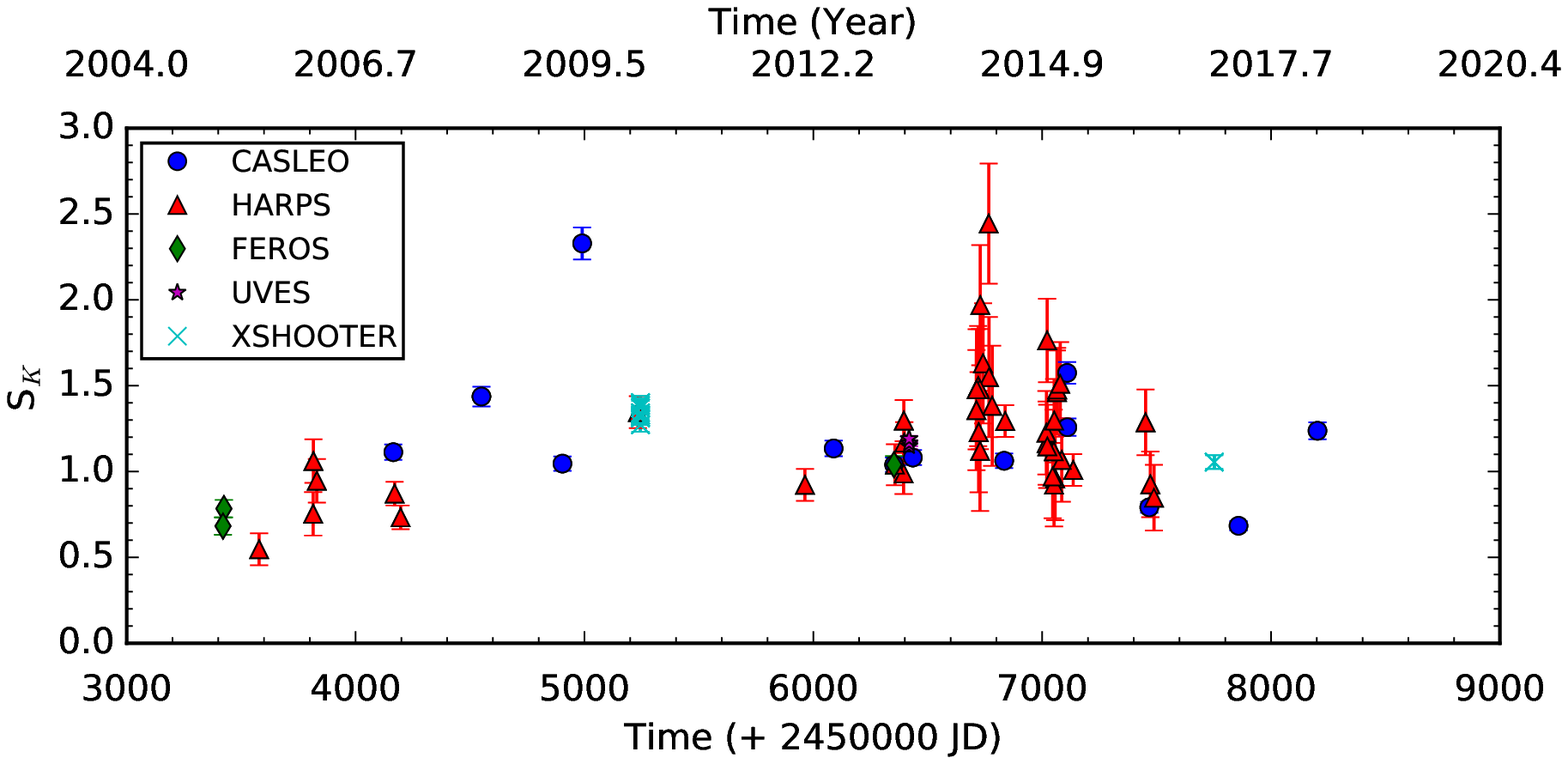}
       \caption{$S_K$-indexes for Ross 128 derived from CASLEO (\textcolor{blue}{$\bullet$}), HARPS      (\textcolor{red}{$\blacktriangle$}), FEROS (\textcolor[cmyk]{1,0,1,0.4}{$\blacklozenge$}), UVES  (\textcolor{violet}{$\bigstar$}) and XSHOOTER (\textcolor{cyan}{$\times$}) spectra.}%
   \label{st_CHFUX}
\end{figure*}

In Fig. \ref{st_CHFUX} we show the CASLEO $S_K$ time series compiled for Ross 128 between 2006 and 2018, combined with the $S_K$-indexes obtained from HARPS, FEROS, UVES and XSHOOTER spectra between 2004 and 2017.  We intercalibrated the CASLEO, FEROS, UVES and XSHOOTER indexes with those of HARPS that were closest in time. We chose the HARPS series as reference because it is the one with both most observations and a wide extension in time. From the whole spectroscopic data series we obtain a variability of $\sigma_{S(K)}$/$\langle S_K \rangle$ $\sim$ 25\%. 

Following \cite{Ibanez18}, we quantify the typical errors of CASLEO spectra by studying the $S_K$ index of the flat star $\tau$ Ceti (HD 10700). In this new analysis, we obtained a dispersion of 1$\sigma_{S(K)}$/$\langle S_K \rangle$ $\sim$ 4 \% coincident with the Mount-Wilson index dispersion.

The error bars of HARPS, FEROS, UVES and XSHOOTER were obtained in the following way: we group the indexes in monthly bins and calculate the standard deviation of each bin. For those time intervals with only one index in a month, we adopted the typical RMS dispersion of the  bins. In the case of HARPS observations, each point represents the average of 2 to 4 nights. The error bars give the idea of the monthly  variability of Ross 128, which is larger during its maximum in 2014.

\begin{figure}
   \centering
      \includegraphics[width=9cm]{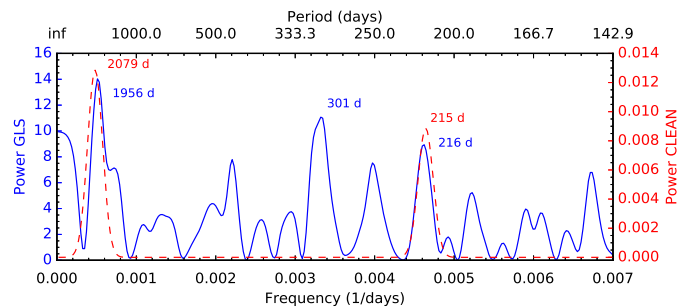}
   \caption{ GLS (blue solid line) and CLEAN (red dashed line) periodograms for the whole $S_K$ time series of Ross 128. There are three prominent peaks for the GLS periodogram: $(1956 \pm 98)$ days, $(301 \pm 4)$ days and $(216 \pm 2)$ days with FAPs of $2\times10^{-6}$, $1\times10^{-4}$ and $2\times10^{-3}$, respectively.}%
   \label{per_gls-clean}
    \end{figure}

To the whole time series, we computed the Generalized Lomb-Scargle (GLS) periodogram \citep{Zechmeister09} to search for long-term activity cycles (Fig.\ref{per_gls-clean}). The periodogram shows two significant peaks at $P = (1956 \pm 98)$ days ($\sim$ 5.4 years) and $P = (301 \pm 4)$ days ($\sim$ 0.8 year) with very good FAPs of  2$\times10 ^{-6}$ and 1$\times10 ^{-4}$, respectively. The error of the period detected depends on the finite frequency resolution of the periodogram $\delta\nu$ as given by Eq.(2) in \cite{Lamm04}, $\delta P=\frac{\delta\nu P^2}{2}$. 

To investigate whether the 301-days period is due to sampling, we implemented the CLEAN deconvolution algorithm described in \cite{Roberts87}. In Fig. \ref{per_gls-clean} we also show the comparison between GLS (blue) and CLEAN (red) periodograms obtained for the whole data. The prominent peak of $\sim$5 yr is present in both the GLS and CLEAN periodograms, but the 301-days period is absent in the CLEAN power spectrum. Therefore this peak is most probably an alias induced by the spectral window function. On the other hand, a 216-days peak  is still present in both periodograms, with a FAP of $2\times10^{-3}$ for the GLS method. As this period could be an alias of the 121-days rotation period reported by \cite{Bonfils18}, we  estimated the stellar $P_{rot}$ from our activity registry. We first restricted the time series only to spectra with a good signal-to-noise ratio on Ca \scriptsize{II}\normalsize\- H line. From these data,  we obtained a mean Mount Wilson index $\langle S \rangle = 1.211 \pm 0.230$ and a Ca \scriptsize{II}\normalsize\- emission level of $\log (R'_{HK}) = -5.621$, both in agreement with the values reported by \cite{Astudillo17}. Using  Eq. (12) of  Astudillo-Defru's work, we obtained a rotation period of $\sim$109 days, which is nearly half 216 days. \textbf{Although, the 109-day period is slightly different from \citeauthor{Bonfils18}'s period, both are consistent withe level of activity of the star. A more exhaustive observation of Ross 128 should be done to determine a reliable rotation period.}

\begin{figure}
   \centering
      \includegraphics[width=9cm]{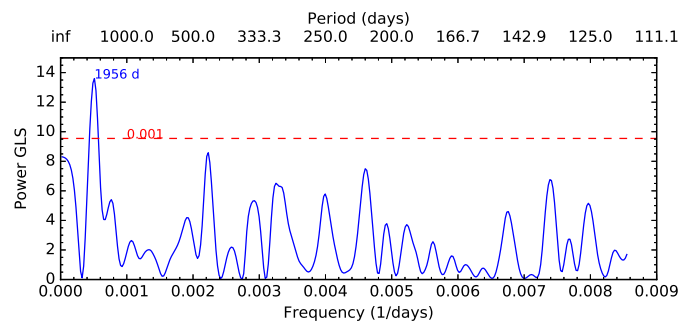}
   \caption{GLS periodogram after subtracting the $\sim$109-days period..}%
   \label{residuos_109}
    \end{figure}

We subtracted from our $S_K$ series the modulation with the 109-days rotation period that we found from the previous analysis. For the new residual series, we compute the GLS periodogram, and we found only one prominent peak at $\sim$5 yr (Fig. \ref{residuos_109}). \textbf{Moreover, if from our $S_K$-series we subtract the 121-days rotation period reported by \cite{Bonfils18}, the analysis of the residual series is similar to that obtained in Fig. \ref{residuos_109}.} Therefore we discard the 216-days peak as an activity signal and we conclude that it should be related to rotational modulation. 

\subsection{Sodium activity indicators }

\begin{figure}
\centering
    \includegraphics[width=9cm]{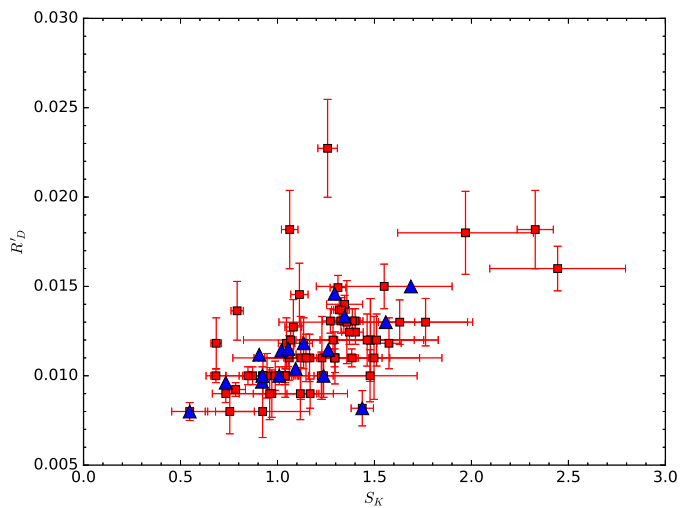}
       \caption{Simultaneous measurements of the Na \scriptsize{I}\normalsize\- index $R'_D$ and the $S_K$-index. The blue triangles indicate the spectroscopic data grouped every 109 days.}%
   \label{comparo_indices}
\end{figure}

Given the spectral range of most of the spectra employed in this analysis, we are able to explore  activity indicators. In particular, the Na \scriptsize{I}\normalsize\- D lines provide information about the conditions in the middle-to-lower chromosphere \citep{Diaz07b}. For early M stars, the sodium doublet profiles present extensive wings (more than 20 Å from the line centre) and their depth is around 11\% of the nearby continuum \citep{Houdebine09}.

\cite{Houdebine09} studied the sodium spectral lines for M1 stars. They found a good correlation between the Na \scriptsize{I}\normalsize\- D1 and D2 lines cores fluxes showing that the optical depths decrease with an increasing activity level, opposite to the Ca \scriptsize{II}\normalsize\- lines. 

M dwarf stars are usually divided between inactive dM stars, with the H$\alpha$ line in absorption, and active dMe stars, with H$\alpha$ in emission. Using high-resolution HARPS spectra, \cite{GomesdaSilva12} studied the sodium lines in a sample of low-activity stars with spectral range from M0 to M3.5 and the late dMe stars Proxima Cen and Barnard star. They found that 70\% of their sample shows a strong and positive correlation between the Mount Wilson and the Sodium indexes. 

In this work, we explore the correlation between the Ca  \scriptsize{II}\normalsize\ and the Na \scriptsize{I}\normalsize\- lines with only those spectra which included both features. We computed the Sodium  $R'_D$-index defined by \cite{GomesdaSilva12}. In Fig. \ref{comparo_indices}, we plot simultaneous measurements of the $R'_D$-index versus the $S_K$-index. We found that both indexes are correlated with a Pearson's correlation coefficient of $R = 0.56$. To remove short-term variations due to rotational modulation, we grouped the spectroscopic data every 109 days and we found that it remains positive with a Pearson's coefficient of $R = 0.65$ for the binned data (blue triangles in Fig. \ref{comparo_indices}).
Finally, we built a time series with the $R'_D$ binned indexes and we found a period of $P_{Na} = (2151 \pm 285)$ days, similar to the one detected with the $S_K$ series but with a larger FAP $=$ 0.14, probably related to the long-term activity of Ross 128.

\section{Conclusions and summary}

The stellar dynamo in fully-convective stars has been scarcely explored.  In this work, we studied the long-term chromospheric activity of the slow-rotator dM4 star Ross 128.  We derived the $S_K$ activity index from the Ca \scriptsize{II}\normalsize\- K line and we built for the first time a long-time series employing CASLEO, HARPS, FEROS, UVES and XSHOOTER spectra obtained over a span of 14 years. 
For the whole time series we detected a significant $\sim 5.4$ yr period with both the GLS and CLEAN periodograms. The fact that we obtained this period with both formalisms reinforces the significance of our detection. Furthermore, since most of the spectra collected of Ross 128 cover a wide wavelength range, we were able to study the magnetic activity in the middle chromosphere using the Na \scriptsize{I}\normalsize\- index. We also detected an activity cycle of $P_{Na} \sim 5.8$ yr with a FAP of 0.14. Moreover, we found a good correlation between the Na \scriptsize{I}\normalsize\- and the $S_K$ indexes, which remains positive during the whole cycle. 

Several cycles have been reported in  late-type stars (e.g., \citealt{Cincunegui07, Buccino11, Metcalfe13, Ibanez18}). However, all these stars belong to the saturation regime in the $L_X/L_{bol}$ vs. $P_{rot}/\tau$ diagram  \citep{Wright11, Wright18} and the $ \log(R'_{HK})$ vs. $P_{rot}$ diagram  \citep{Astudillo17}). Until now, only one activity cycle of a slow-rotator star was reported in the literature: the active M5.5 star Proxima Centauri (\citealt{Cincunegui07, Wargelin17}). Despite presenting a rotation period of $\sim$80 days, Proxima Cen shows extremely frequent flares and even superflares \citep{Davenport16}. Therefore, due to its activity level, this slow-rotator may not belong to the unsaturated regime \citep{Astudillo17}. 

From our registry of activity, we obtained a mean activity index $\log R'_{HK} = -5.621$ for Ross 128. If we consider the 121-days rotation period reported by \cite{Bonfils18}, the $\log R'_{HK} $ value place this fully-convective M dwarf in the non-saturated regime. From X-ray observations, $L_X/L_{bol} \sim 10^{-5}$ for Ross 128 (\citealt{Fleming95, Malo14, Stelzer16}), which implies that this slow-rotator is also outside the saturation regime in the  $L_X/L_{bol}$ vs $P_{rot}$ diagram. On the other hand, in the optical range \cite{Newton17} reported a relation between $L_{H\alpha}/L_{bol} $ and the Rossby number placing Ross 128 in the power-law decay regime. Moreover, they found a clear mass-dependent rotation period for inactive stars and estimated  a rotation period $\sim 103$ days for this star. This value is in agreement within the statistical error with the one reported by \cite{Bonfils18} from ASAS photometry  and the one that we estimated from the $log(R'_{HK})$-index.

The activity cycle detected in this work for Ross 128 is in agreement with \cite{Wright16}, who sustain that a solar-dynamo could operate in the fully-convective stars in the non-saturated regime. In this way, Ross 128 becomes the one of the few slow-rotator stars of its class outside the saturation regime to present a stellar activity cycle.

Likewise, M dwarfs are ideal targets to search for terrestrial planets in the habitable zone. However, stars such as Proxima Centauri (dM5.5e) and TRAPPIST-1 (dM8e), show frequent high-energy flares that could impact on the planetary atmospheres and constrain the habitability of their orbiting planets \citep{Buccino06, Buccino07, Hawley14}. Recently, \cite{Gunther19} reported that M4-M6 dwarfs are the most common flare stars with a 30\% of the total M stars showing flares.
In addition dM stars (without observable H$\alpha$ emission activity) are the most abundant in our galaxy. They present flare frequencies much lower than the active ones of the same spectral type and few of these flares can release high energy \citep{Hawley14}. Furthermore, \cite{Yang17} found that the flare activity ($L_{flare}/L_{bol}$) increases as rotation period decreases, to reach an activity saturation around $\sim 5\times10^{-5}$ at periods near $\sim 5$ days. They also suggest that $P_{rot} \sim 10$ days  or $L_{flare}/L_{bol} \sim 10^{-6}$ could be a boundary between active and  inactive stars. However, above the rotation period of 10 days, active stars are mixed with inactive ones. Although flares were also detected in inactive stars, the flare rate and flare energy of inactive and fully-convective stars with rotation periods greater than 100 days is not well characterized. As Ross 128 presents signs of long-term activity, it would be an excellent target for the Transiting Exoplanet Survey Satellite (TESS) to shed light on its flare rate.




\bibliographystyle{aa}
\small
\bibliography{biblio}

\end{document}